\documentclass[twocolumn,showpacs,preprintnumbers,amsmath,amssymb]{revtex4}
\usepackage{graphicx}% Include figure files
\usepackage{dcolumn}% Align table columns on decimal point
\usepackage{bm}% bold math
\begin{document}
\title{Effects of Community Structure on Search and Ranking in Information Networks}
\author{Huafeng Xie$^{1,3}$, Koon-Kiu Yan$^{2,3}$, Sergei Maslov$^{3}$
\footnote{To whom the correspondence should be addressed: maslov@bnl.gov}}
\affiliation{
$^1$New Media Lab, The Graduate Center, CUNY
New York, NY 10016, USA\\
$^2$Department of Physics and Astronomy, Stony Brook University, \\
Stony Brook, New York, 11794, USA\\
$^3$Department of Physics, Brookhaven National Laboratory,
Upton, New York 11973,  USA}
%\altaffiliation{Department of Physics, Brookhaven National Laboratory,
%Upton, New York 11973,  USA}
%\author{Koon-Kiu Yan}
%\affiliation{Department of Physics, SUNY at Stony Brook, New York, 11794}
%\altaffiliation{Department of Physics, Brookhaven National Laboratory,
%Upton, New York 11973,  USA}
%\author{Sergei Maslov}
%\email{maslov@bnl.gov}
%\affiliation{Department of Physics, Brookhaven National Laboratory,
%Upton, New York 11973,  USA}
\date{\today}
\begin{abstract}
The World-Wide Web (WWW) is characterized by a strong community
structure in which communities of webpages (e.g. those sharing a
common keyword) are densely interconnected by hyperlinks. We study how such network
architecture affects the average Google ranking of individual
webpages in the comunity. It is shown that the Google rank of community
webpages could either increase or decrease with the density of
inter-community links depending on the exact balance between
average in- and out-degrees in the community. The magnitude of
this effect is described by a simple analytical formula and
subsequently verified by numerical simulations of random
scale-free networks with a desired level of the community
structure. A new algorithm allowing for generation of such
networks is proposed and studied. The number of inter-community
links in such networks is controlled by a temperature-like
parameter with the strongest community structure realized in
``low-temperature'' networks.
%
%To generate artificial scale-free networks with a desired level of
%community structure we developed a version of the Metropolis
%rewiring algorithm first proposed in \cite{maslov physicaA 2004}.
%The density of links connecting the set of pre-selected community
%nodes to each other is controlled by a temperature-like parameter
%of the this algorithm so that the ``low-temperature'' networks
%have the highest possible density of inter-community links.
%
\end{abstract}
\maketitle
\begin{figure}
[htbp]
\centering
\includegraphics*[width=2in]{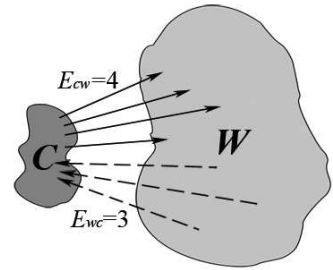}
\caption{The illustration of hyperlink connections between the
community $C$ and the outside world $W$. $E_{cw}$ and $E_{wc}$ are
numbers of links from the community to the outside world and from
the outside world to the community, respectively.} \label{cw_ill}
\end{figure}

The World Wide Web (WWW) -- a very large ($\sim 10^{10}$ nodes)
network consisting of webpages connected by hyperlinks -- presents
a challenge for the efficient information retrieval and ranking.
Apart from the contents of webpages, the topology of the network
itself can be a rich source of information about their relative
importance and relevance to the search query. It is the effective
utilization of this topological information \cite{PageBrin} which
advanced the Google search engine to its present position of the
most popular tool on the WWW and a profitable company with a
current market capitalization around \$30 billion. To rank the
importance of webpages Google simulates the behavior of a large
number of ``random surfers" who just follow a randomly selected
hyperlink on each page they visit. The number of hits a given page
gets in the course of such simulated process determines its
ranking. It is intuitively clear that the larger is the number of
hyperlinks pointing to a given webpage (its in-degree in the
network) the higher are the chances of a random surfer to click on
one of them and, therefore, the higher would be the resulting
Google rank of this webpage. However, the algorithm goes beyond
just ranking nodes based on their in-degrees. Indeed, the traffic
directed to a given webpage along a particular incoming hyperlink
is proportional to the popularity of the webpage containing this
link. Therefore, the Google rank of a node is given by the
weighted in-degree where the weight of each neighboring webpage
reflects its importance and is determined self-consistently.
%\par
The WWW is a very heterogeneous collection of webpages which can
be grouped based on their textual contents, language in which they
are written, the Internet Service Provider (ISP) where they are
hosted, etc. Therefore, it should come as no surprise that the WWW
has a strong community structure \cite{community} in which similar
pages are more likely to contain hyperlinks to each other than to
the outside world. Formally a web community can be defined as a
collection of webpages characterized by a higher than average
density of links connecting them to each other. In this letter we
are going to address the question: how the community structure
affects the Google rank of webpages inside the community. One
might naively expect that the community structure always boosts
the Google rank of its webpages as it tends to ``trap'' the random
surfer inside the community for a longer time. However, it turned
out that it is not generally true. In fact the Google rank of
community webpages could either increase or decrease with the
density of inter-community links depending on the exact balance
between average in- and out-degrees in the community. In the heart
of the Google search engine lies the PageRank algorithm
determining the global ``importance'' of every web page based on
the link structure of the WWW network around it. While the details
of the algorithm have undoubtedly changed since its introduction
in 1997, the central ``random surfer'' idea first described in
\cite{PageBrin} remained essentially the same. To a physicist the
algorithm behind the PageRank just simulates an auxiliary
diffusion process taking place on the network in question.
Similar diffusion algorithms have been recently applied to study
citation and metabolic networks \cite{Peterson} and the
modularity of the Internet on the ``hardware level'' represented
by an undirected network of interconnections between Autonomous
Systems \cite{Maslov03}. A large number of random walkers are
initially randomly distributed on the network and are allowed to
move along its directed links. As in principle some nodes in the
network could have zero out-degree but non-zero in-degree and
would thus ``trap'' random walkers, the authors of the algorithm
introduced a finite probability $\alpha$ for a surfer to randomly
select a page in the network and directly jump there without
following any hyperlinks. This leaves the probability $1-\alpha$
for a surfer to randomly select and follow one of the hyperlinks
of the current webpage. According to \cite{PageRank} the original
PageRank algorithm used $\alpha=0.15$.
%
%random walker to move to one of the neighboring
%nodes along a randomly selected outgoing link. PageRank value of
%node $i$, denoted as $x_i$, is defined as the number of random
%walkers present in node $i$ in the limiting stationary distribution
%of the described diffusion process. Here, the factor $\alpha$ plays
%an important role in the picture. Without the possibility of random
%jumping, all random walkers will finally be trapped in nodes with
%zero outgoing link.
%
The PageRank then simulates this diffusion process until it
converges to a stationary distribution. The Google rank (PageRank)
$G(i)$ of a node $i$ is proportional to the number of random
walkers at this node in such steady state. We chose to normalize
it so that $\sum_{i} G(i)=1$ but in general the normalization
factor does not matter as ranking relies on relative values of
$G(i)$ for different webpages. When one enters a search keyword
such as e.g. ``statistical physics'' on the Google website the
search engine first localizes the subset of webpages containing
this keyword and then simply presents them in the descending order
based on their PageRank values.
%
%
%Let us write down an expression for PageRank values explicitly. For
%any node $i$, let $V_i$ be the set of nodes pointing to $i$, and
%$k_i$ be the number of outgoing links from node $i$, then
%
The main equation determining the PageRank values $G(i)$ for all
webpages in the WWW is
\begin{equation}
G(i)=\alpha+\sum_{j \to i} (1-\alpha) \frac{G(j)}{K_{out}(j)}  .
\label{PR}
\end{equation}
Here $K_{out}(j)$ denotes the the number of hyperlinks (the
out-degree) in the node $j$ and the summation goes over all nodes
$j$ that have a hyperlink pointing to the node $i$. In the matrix
formalism the PageRank values are given by the components of the
principal eigenvector of an asymmetric positive matrix related to
the adjacency matrix of the network. Such eigenvector could be
easily found using a simple iterative algorithm \cite{PageRank}.
The fast convergence of this algorithm is ensured by the fact that
the adjacency matrix of the network is sparse.
%
%It is worth to point out the similarity and
%difference between PageRank value of a node and the in-degree of the
%node. In general, the in-degree of a node has a positive effect to
%the PageRank value of the node. However, PageRank value takes into
%account the ``importance" of the neighboring nodes. According to
%equation (\ref{PR}), neighboring node with a high PageRank value and
%a low out-degree has greater contribution to the PageRank value of a
%node.
We first consider the effect of the community structure on Google
ranking in the simplest and most physically transparent case of
$\alpha=0$. In order for the algorithm to properly converge in
this case we need to assume that $K_{out}(i)>0$ for all nodes in
the network. Consider a network in which $N_c$ nodes form a
community characterized by higher than average density of edges
linking these nodes to each other. Let $E_{cw}$ denote the total
number of hyperlinks pointing from nodes in the {\it community} to
the outside {\it world}, while $E_{wc}$ - the total number of
hyperlinks pointing in the opposite direction (See Fig.
\ref{cw_ill} for an illustration).
Similarly $E_{cc}$ and $E_{ww}$ denote the total number of links
connecting nodes within the community and, respectively, the
outside world. The total number of hyperlinks pointing to nodes
inside the community is given by $E_{cc}+E_{wc}=N_c \langle K_{in}
\rangle_c$ where $\langle K_{in} \rangle_c$ is the average
in-degree of community nodes. Similarly,
$E_{cc}+E_{cw}=N_c \langle K_{out} \rangle_c$, where
$\langle K_{out} \rangle_c$ is the average out-degree in the
community, gives the total number of hyperlinks originating on
community nodes.
The Google rank is computed in the steady state of the
diffusion process where the average number of random surfers
currently visiting any given webpage does not change with time. This means
that the total current of surfers $J_{cw}$ leaving the community for the
outside world must be precisely balanced by the
current $J_{wc}$ entering the community during the same time interval.
%
%$G_{c}$ and $G_{w}$ are the average Google rank of the community
%members and the rest of the world; $\langle K_{out} \rangle_{c}$
%and $\langle K_{out} \rangle_{w}$ are the number of the average
%outgoing links of the community members and the rest of the world,
%
%\begin{figure}
%\includegraphics{fig_1A}% Here is how to import EPS art
%\caption{\label{fig:epsart} Illustration of the links between the
%community and the world.}
%\end{figure}
%
Let $G_{c}=\langle G(i) \rangle_{i \in C}$
denote the average Google rank inside the community
given by the average number of random surfers on its nodes.
If edges pointing away from the community to the outside
world start at an unbiased selection of nodes in the community the average current
flowing along any of those edges would be given by $G_{c}/\langle
K_{out} \rangle_{c}$ while the total current leaving the
community $J_{cw} =E_{cw} G_{c}/\langle K_{out}
\rangle_{c}$. Similar analysis gives
$J_{wc} =E_{wc} G_{w}/\langle K_{out} \rangle_{w}$,
where $\langle K_{out}\rangle_{w}$
is the average out-degrees of nodes in the world outside the
community.
%
%So we have:
%$G_{c} {E_{cw}\over \langle K_{out} \rangle_{c}}= G_{w}
%{E_{wc}\over \langle K_{out} \rangle_{w}}$
%
%A simple derivation gives:
Balancing these two currents one gets:
\begin{equation}
\frac{G_{c}}{G_{w}}=  \frac{E_{wc}}{E_{cw}}\cdot \frac{\langle K_{out}
\rangle_{c}}{\langle K_{out} \rangle_{w}} \qquad .
\label{eq_1}
\end{equation}
The Eq. \ref{eq_1} is based on the ``mean-field'' assumption that average
values of the Google rank and the out-degree on those community nodes that actually
send links to the outside world are equal to their overall average values
inside the community \cite{comment}.
%it may be good to use subscript to denote average like G_{c} and G(i) to
%denote rank of individual node
%If the community size is large enough (while it's still small
%comparing to the world), one can expect to find that $\langle
%K_{out} \rangle_{c}$ and $\langle K_{out} \rangle_{w}$ are close.
%So the ratio of $E_{cw}$ and $E_{wc}$ is the real determinative
%factor the Google rank of a community.
It is tempting to assume that higher than average density of
hyperlinks connecting nodes in the community
is beneficial for the Google rank of its
nodes as it ``traps'' random surfers to spend more time within the community.
It turned out that this naive argument is not necessarily true. In fact one is equally likely
to observe an opposite effect: an excess of intra-community links
%while keeping in- and out-degrees of its nodes fixed
could lead to a lower than average Google rank of its nodes.
%
%To clarify the ambiguity, we need to
%separate the effects of different aspects of community structure.
%First we denote the total number of incoming links of nodes inside
%the community as $\langle K_{in}\rangle_{c}N_{c}$, where $\langle
%K_{in}\rangle_{c}$ is the average in degree of the nodes inside
%the community and $N_{c}$ is the total number of nodes inside the
%community.
%For the nodes inside the community, the incoming links
%originate either from the nodes inside the community or the nodes
%of the world. The total number of the former kind of links is
%$E_{cc}$ and total number of the later kind of links is $E_{wc}$
%according to our definition of $E_{cc}$ and $E_{wc}$. So $E_{wc}$
%can be denoted by $\langle K_{in}\rangle_{c}N_{c}-E_{cc}$.
%Similarly we denote $E_{cw}$ as $\langle
%K_{out}\rangle_{c}N_{c}-E_{cc}$.
To see it explicitly one should replace $E_{wc}$ and $E_{cw}$ in Eq. \ref{eq_1}
with identical expressions $\langle K_{in}\rangle_{c}N_{c}-E_{cc}$
and $\langle K_{out}\rangle_{c}N_{c}-E_{cc}$ respectively:
\begin{equation}
\frac{G_{c}}{G_{w}}=  \bigg(\frac{\langle
K_{in}\rangle_{c}N_{c}-E_{cc}}{
{\langle K_{out}\rangle_{c}N_{c}-E_{cc}}}\bigg)\cdot \frac{\langle
K_{out}\rangle_{c}}{\langle K_{out} \rangle_{w}} \qquad .
\label{maineq_a}
\end{equation}
From this equation it follows that enhancing the community structure (increasing
$E_{cc}$) while keeping other parameters such as $\langle K_{in}
\rangle_{c}$,$\langle K_{out} \rangle_{c}$ and $\langle K_{out}
\rangle_{w}$ fixed can be both good and bad for the average Google
rank of the community webpages. It depends on $\langle
K_{in} \rangle_{c}/ \langle K_{out}\rangle_{c}$ -- the ratio between average
in- and out-degrees of community nodes. If the ratio is less than
1 the increase in $E_{cc}$ leads to a further decrease of
$G_{c}/G_{w}$ below one. If the community constitutes just a small
fraction of the whole network one could safely assume that $G_{w}$ remains approximately
constant so that the average Google rank of the community, $G_c$,
has to decrease. Similarly if the ratio is larger than 1,  $G_c$
grows with the number of inter-community links $E_{cc}$ (see Fig. 2
for an illustration of both cases).
\begin{figure}[htbp]
\centering
\includegraphics*[width=2.5in]{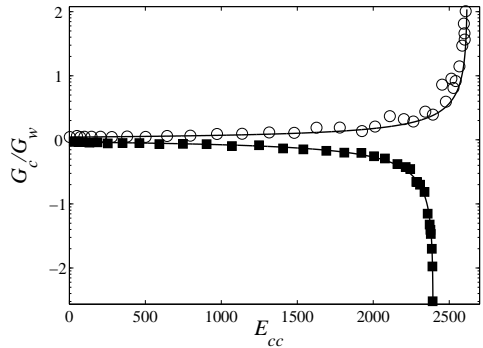}
\caption{The ratio of average Google ranks in the community
and the outside world $G_c/G_w$ as a function of $E_{cc}$ --
the number of intra-community links
-- in two series of model networks with varying degree of community
structure. Open circles correspond to a beneficial effect of the
community structure on Google ranking in a scale-free network with
$\langle K_{out}\rangle_{c}=5.24 <  \langle K_{in}\rangle_{c}=5.9$.
On the other hand, filled squares show a detrimental effect in another series of networks
where $\langle K_{out}\rangle_{c}=5.6 > \langle
K_{in}\rangle_{c}=4.8$.
Solid lines are fits with the Eq. \ref{maineq_a} with a
given set of parameters for each of the networks.
All networks with $10,000$ nodes have a community of $500$ nodes
were generated by the Metropolis rewiring algorithm described
later on in the text.}
\end{figure}
The real-life Google algorithm uses a non-zero value of $\alpha \simeq 0.15$.
In this case one needs to consider the contribution to currents
$J_{cw}$  and $J_{wc}$ due to surfers' random
jumps that do not follow the existing hyperlinks.
%According to the Google algorithm, a random walker has
%probability $1-\alpha$ to follow one of the node's outgoing links
%and probability $\alpha$ to jump to a random node.
The total number of random walkers residing on the nodes inside the community is
$G_{c} N_{c}$ and the probability of them to randomly jump to a
node in the outside world is
$N_{w}/(N_{c}+N_{w})$. So the contribution to the outgoing current
due to such jumps is given by $\alpha G_{c}N_{c}N_{w}/(N_{c}+N_{w})$
which for $N_{c} \ll N_{w}$ can be simplified as $\alpha
G_{c}N_{c}$. The total outgoing current then can then be written as
$J_{cw} =(1-\alpha) G_{c} E_{cw} / \langle K_{out}\rangle_{c}
+\alpha G_{c} N_{c}$.
%For $N_{c} \ll N_{w}$ one has
%$J_{cw} \simeq  E_{cw}{G_{c}\over\langle K_{out} \rangle_{c}}(1-\alpha)+G_{c}N_{c}\alpha$.
Similarly the incoming current $J_{wc}$ is given by $(1-\alpha)G_{w}
E_{wc}/\langle K_{out} \rangle_{w} +\alpha G_{w} N_{c}$.
The Eq. \ref{eq_1} remains valid for $\alpha>0$ if one replaces
$E_{wc}$ and $E_{cw}$ with ``effective'' numbers of edges $E_{wc}^*$ and $E_{cw}^*$
given by
\begin{eqnarray}
E_{cw}^*=E_{cw}(1-\alpha)+N_{c}\langle K_{out}\rangle_{c}\alpha \qquad
; \nonumber \\
E_{wc}^*=E_{wc}(1-\alpha)+N_{c}\langle K_{out}\rangle_{w}\alpha
\qquad .
\end{eqnarray}
These effective numbers take into account contributions to both currents due to
random jumps.
% as well as that following the hyperlinks.
%
%
%we obtain $J_{cw}^{*} = E_{cw}^{*}{G_{c}\over \langle K_{out}
%\rangle_{c}} $ and $J_{wc}^{*} = E_{wc}^{*}{G_{w}\over \langle
%K_{out} \rangle_{w}}$. When $\alpha>0$, Eq.\ref{eq_1} can be rewritten as $
%{G_{c} \over G_{w}}=  {E_{wc}^{*}\over E_{cw}^{*}}{\langle K_{out}
%\rangle_{c}\over \langle K_{out} \rangle_{w}}$ and if one assumes
%that $\langle K_{out} \rangle_{c}$ and $\langle K_{out}
%\rangle_{w}$ are proximately the same, this equation can be
%further simplified to:
%\begin{equation}
%{G_{c} \over G_{w}}=  {E_{wc}^{*}\over E_{cw}^{*}}
%\label{maineq_simple}
%\end{equation}
For a numerical test of the validity of our analytical results we
generated an ensemble of directed networks with scale-free
distributions of in- and out-degrees: $P(K_{in})\sim
K_{in}^{-2.1}$ and $P(K_{out})\sim K_{out}^{-2.5}$
correspondingly. The exponents were selected to be identical to
their values in the actual WWW network
\cite{community,WWW_exponents}. The community structure in those
networks was artificially created using the Metropolis rewiring
algorithm described in the next section. As a result a
pre-selected group of $N_c$ nodes formed an artificial community
with the exact number of intra-community links controlled by the
parameters of our simulation. The Fig. \ref{fig_test_eq_1} shows
the results of a numerical test of Eq. \ref{eq_1} in those model
networks.
%and tested Eq.\ref{eq_1} and Eq.\ref{maineq_simple}.
%
\begin{figure}[htbp]
\centering
\includegraphics*[width=2.5in]{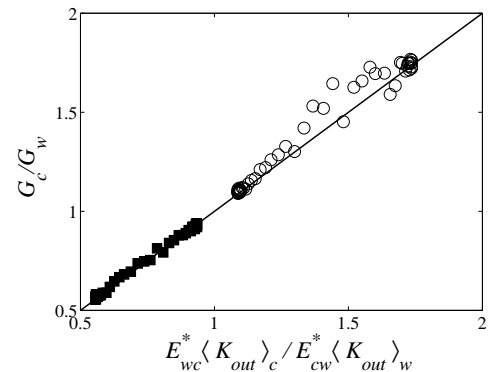}
\caption{The ratio of average Google ranks in the community
and the outside world $G_c/G_w$ as a function of the ratio of
effective numbers of links $E^*_{wc}/E^*_{cw}$. As predicted by
the Eq. \ref{eq_1} these two ratios are basically equal to each
other. Different symbols correspond to series of networks described
in Fig. 2}
\label{fig_test_eq_1}
\end{figure}
% Fig.2 $G_{c}$ vs $E_{nc}/E_{cn}$ and in the insert how using
% $E^*_{nc}/E^*_{cn}$ makes it "trivial".
%To check how the theory works, we came up with
%An algorithm used in Figs. 2,3
%allows us
%
%The main subject of this work is how a community structure in
%networks in general and in scale-free directed networks in
%particular affects the rest of their topological properties.
For numerical studies of networks with a community structure one
needs an efficient algorithm to generate them. In this work we
propose a version of the Metropolis
%\cite{metropolis}
random rewiring algorithm introduced earlier in
\cite{maslov_physicaA_2004}.
%It is sometimes desirable to generate a version of the network
%to generate a community structure on any given
%subset of nodes would form a ``community" manifested by
%a higher density of links connecting these nodes to each other. In
%addition, our algorithm allows us to generate an ``anti-community"
%structure as well in which the density of connections between
%selected nodes is below average.
The algorithm starts from a ``seed'' network with the
desired (scale-free in our case) distributions of in- and out-degrees.
Such a seed network can be created e.g. using a stub reconnection
procedure described in \cite{newman_stub}. The heart of our algorithm is the
local rewiring (edge switching) step
%that randomizes a directed network yet
which strictly conserves separately the in- and out-degrees of every node
involved \cite{rewire}.
The only parameters of the Metropolis part of our algorithm are
an auxiliary Hamiltonian (energy function) $H=-E_{cc}$ defined as
the negative of the number of intra-community links
and the inverse temperature $\beta$.
%
%Beginning from
%an arbitrary directed network with a prescribed degree
%distribution, a small subset of nodes are presumably selected as
%the community and we will focus on the density of links between
%them. Successive application of the algorithm generates an
%ensemble of networks with the same degree distribution, hence the
%same density of links. However, the density of links within the
%chosen nodes varies among the ensemble elements.
%
%In order to characterize the ensemble generated by the local
%rewiring algorithm, we modify the original algorithm to a
%canonical Metropolis algorithm, with a Hamiltonian $H$ defined as
%the negative of the number of links connecting the nodes belonging
%to the community.
%
The steps of the algorithm are as follows:
%\begin{enumerate}
%
%\item
1) Randomly pick two links, say A$\to$B and C$\to$D;
%from node $A$ to
%node $B$ and from node $C$ to node $D$;
%
%\item
2) Attempt to rewire them (switch their neighbors) to A$\to$D and
C$\to$B.
%so that new links
%would be
%connect the node $A$ to the
%node $D$ and the node $C$ to the node $B$.
If at least one of these two
new links already exists in the network, abort this step
and go back to step 1;
%
%\item
3) If the rewiring step decreases the Hamiltonian $H$ it is always accepted, while if it
increases the Hamiltonian by $\Delta H$ it is accepted only with probability
$\exp(-\beta \Delta H)$. If the rewiring step is
rejected on steps 2 or 3, the network is returned to the
original configuration A$\to$B and C$\to$D;
%
%\item
4) Repeat the above steps until the number of links inside the community
$E_{cc}$ reaches a steady state value.
%
%\end{enumerate}
%
The reciprocal temperature $\beta$ thus indirectly determines the
number of links within the community $E_{cc}(\beta)$ so that an ordinary
random (scale-free) network without any community structure is
realized at an ``infinite temperature''
($\beta=0$), while the algorithm run at zero temperature ($\beta=\infty$)
produces a network with the largest possible number of links within
the community. One could also invert the sign in the definition of the
Hamiltonian $H=E_{cc}$. Formally this can be thought of as running the
algorithm with the original Hamiltonian but a negative inverse temperature
$\beta<0$. Large
negative values of $\beta$ generate networks with an
anti-community structure in which the number of intra-community
links is lower than in a random network. The relation between
$E_{cc}$ and $\beta$ for both positive and negative values of
$\beta$ is shown in Fig. \ref{fig_beta_Ecc}.
\begin{figure}
[htbp]
\centering
\includegraphics*[width=2.5in]{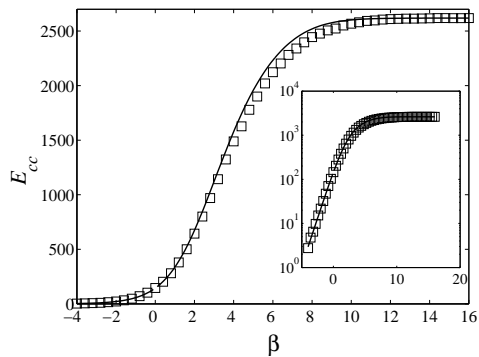}
\caption{The number of intra-community links $E_{cc}$ in
networks generated by the rewiring algorithm as a function of the
inverse temperature $\beta$. Negative values of $\beta$ correspond to networks with
anti-community structure and are generated by changing the sign in front
of the Hamiltonian $H$. The solid line is the fit with the
analytical expression obtained by solving the Eq. \ref{eq_balance}
for $E_{cc}$. The inset shows the same plot with a logarithmic scale of the
Y-axis.}
\label{fig_beta_Ecc}
\end{figure}
% given by the
%smaller of the two numbers $\sum_{j \in C} K_{out}(j)$ and $\sum_{j \in C} K_{in}(j)$.
%, the higher is
%$\beta$, the smaller is the chance that the rewiring steps decreases $E_{cc}$
%gets accepted. So the number of links between nodes inside the
%community becomes larger than it is by pure chance. On the other
%hand, if we allow a negative $\beta$, rewiring that increases $H$(decreases $E_{cc}$)
%always get accepted, while one that decreases $H$(increases $E_{cc}$)
%is accepted with probability $\mbox{exp}(-\beta \Delta H)$.This enable us to
%have an ensemble of networks with anti-community structure, which means that the link
%density within the community is smaller than it is by pure chance.
To analytically derive the relation between $E_{cc}$ and the reciprocal
temperature $\beta$, we consider the detailed balance in the
steady state of the rewiring procedure, in which the probabilities
of an increase and a decrease in $E_{cc}$ must be equal to
each other. $E_{cc}$ is increased by 1 if the links picked
at a given step of the rewiring algorithm are C$\to$W and W$\to$C (here C stands
for any node inside the community and W - in the outside world).
The probability to pick such pair is proportional to
$E_{cw}E_{wc}$. On the other hand, if the selected links are C$\to$C and W$\to$W
the number of links in the community would decrease by one with a
probability $\exp(-\beta)$. All other selections of links do not
change the $E_{cc}$.
%%, at given temperature $T$.
%$E_{cc}$ increases by 1 if a pair of links picked at a given step of rewiring algorithm
%has one link from the community to the outside world and the other is from the outside world to
%the community. The probability of picking the former link is proportional to $E_{cw}$
%and picking the later link is proportional to $E_{wc}$. After
%the rewiring step one new link from community to community and
%another one from the outside world to itself are created.
%On the other hand, when one link of the pair selected before the rewiring
%step is within the
%community and the other is within the outside world, $E_{cc}$
%would decrease 1 but only with probability $\mbox{exp}(-\beta )$ that the
%rewiring move is accepted in the Metropolis algorithm
%(such step would increase the energy function $H$ by 1).
%%(in every rewiring step $H$ can only increase
%%by 1).
%The probability to pick such pair of links is proportional to
%$E_{cc}\cdot E_{ww}$.
%%Then we obtain the equilibrium
The detailed balance equation for the rewiring procedure thus reads:
\begin{equation}
E_{cw}E_{wc}=E_{cc}E_{ww}e^{-\beta}
\label{eq_balance}
\end{equation}
Additional constraints (i) $E_{cc}+E_{wc}=\langle K_{in} \rangle_c N_c$ (the
sum of in-degrees of all nodes within the community), (ii)
$E_{cc}+E_{cw}=\langle K_{out} \rangle_c N_c$ (the sum of out-degrees of all nodes within
the community) and (iii) $E_{cc}+E_{cw}+E_{wc}=E$ (the total
number of edges in the network) plugged into the Eq.
(\ref{eq_balance}) result in a quadratic equation for $E_{cc}$ as a function of
$\langle K_{in} \rangle_c$, $\langle K_{out} \rangle_c$, $E$, and $\beta$ --
the parameters strictly conserved in our rewiring algorithm.
%, given a particular $\beta$, we
%are able to write down analytically the corresponding $E_{cc}$ in
%terms of $D_{in}$, $D_{out}$ and $E$ which are actually conserved
%throughout the rewiring process.
The Fig. \ref{fig_beta_Ecc} compares the analytical expression for $E_{cc}(\beta)$
obtained by solving the Eq. \ref{eq_balance} with numerical simulations for different
values of $\beta$. Clearly, $E_{cc}$ increases with $\beta$ in
general accord with the Eq. \ref{eq_balance}. When $\beta$ is
sufficiently large, $E_{cc}$ exponentially
approaches a limiting value equal to $\max(\langle K_{in} \rangle_c, \langle K_{out} \rangle_c)N_c$
-- the maximal number of links within a community given the set of in- and out-degrees
of its nodes.
%shown in caption that the difference between the limit, say $D_{in}$, and $calculated E_{cc}$ is
%\frac{D_{in}(E-D_{out})}{D_{in}-D_{out}}e^{-\beta}$
%the diff between the limit and the simulation is larger, due to the factor $q/p$.
The deviations between the analytical formula and numerical results visible
for large values of $\beta$ could be attributed to the ``no multiple
edges'' restriction in networks generated by our rewiring
algorithm. As the density of inter-community links increases
with $\beta$ more and more of the rewiring steps leading to an increase of
$E_{cc}$ have to be aborted as the new link they are
attempting to create within a community already exists.
This situation is more appropriately described by the
following equation:
%\begin{equation}
%E_{cw}E_{wc}(1-\frac{E_{cc}}{E})(1-\frac{E_{ww}}{E})
%=E_{cc}E_{ww}(1-\frac{E_{cw}}{E})(1-\frac{E_{wc}}{E})e^{-\beta},
%\qquad ,
%\label{eq_balance_fermion}
%\end{equation}
$E_{cw}E_{wc}(1-E_{cc}/E)(1-E_{ww}/E)
=E_{cc}E_{ww}(1-E_{cw}/E)(1-E_{wc}/E)e^{-\beta}$,
reminiscent of the detailed balance equation in two-fermion scattering
(see also \cite{newman_fermion}).
%
%
%\par One should notice that Eq. \ref{equ_beta} does not take into
%account of the disallowance of multiple edges described in step 2
%of our algorithm. Therefore, the actual probability of changing
%$E_{cc}$, with the absence of pre-existing links presumed, is
%smaller than the one we have just explained. Moreover, as $E_{cc}$
%increases, pre-existing links between the community and the world
%are less common than those within the community. If we denote the
%probability of the absence of pre-existing links within (not
%within) the community be $p$ ($q$), Eq. \ref{equ_beta} should be
%rewritten as $E_{cc}E_{wc}=E_{cc}E_{ww}\,q/p$ where $q/p>1$. We
%believe that this modification explains the small system error
%between numerical simulation and analytical solution in Fig.
%\ref{beta_equ} when $\beta$ is large.

In summary, we investigated how the WWW community structure
affects the Google rank of webpages belonging to a given
community. We have shown
that depending on the balance between average in- and out-degrees of
webpages inside the community the excess density of intra-community
hyperlinks can either boost or decrease the average Google
ranking of its webpages. For numerical studies of scale-free networks with a
community structure we developed a version of the Metropolis
rewiring algorithm first proposed by one of us in \cite{maslov_physicaA_2004}.
This algorithm allows one to generate a random network with a desired
density of intra-community links and a given distribution of in- and out-degrees.
%
%In addition the number of links connecting the set of pre-selected set of
%``community'' nodes in such networks is controlled by a temperature-like parameter
%so that ``low-temperature'' networks have the highest possible
%density of the intra-community links.

Work at Brookhaven National Laboratory was carried
out under Contract No. DE-AC02-98CH10886, Division
of Material Science, U.S. Department of Energy.


\begin{thebibliography}{10}
\bibitem{PageBrin} S. Brin and L. Page,
%""The anatomy of a large-scale hypertextual {Web} search engine"
Computer Networks and ISDN Systems, {\bf 30}, 107 (1998).
%1--7,
%pages = "107--117",
%year = "1998",
%L. Page and S. Brin, in Proceedings of the Seventh International Web Conference
%(Elsevier, Amsterdam,1998).
%
%
\bibitem{community} R. Kumar, P. Raghavan, S. Rajagopalan,  and
A.Tomkins, Computer Networks {\bf 31}, 11 (1999).
%
%author = "Ravi Kumar and Prabhakar Raghavan and Sridhar
%Rajagopalan and Andrew Tomkins",
%    title = "Trawling the {Web} for emerging cyber-communities",
%    journal = "Computer Networks (Amsterdam, Netherlands: 1999)",
%    volume = "31",
%    number = "11--16",
%    pages = "1481--1493",
%    year = "1999",
%
\bibitem{PageRank} L. Page, S. Brin, R. Motwani and T. Winograd,
%"The PageRank Citation Ranking: Bringing Order to the Web"
Stanford Digital Library Technologies Project (1998).
%
\bibitem{Peterson}
%Topological properties of citation and metabolic networks
S. Bilke and C. Peterson
Physical Review E {\bf 64}, 036106 (2001).
%
\bibitem{Maslov03} K. A. Eriksen, I. Simonsen, S. Maslov and K.
Sneppen, Phys. Rev. Lett. \textbf{90}, 148701 (2003).
%
\bibitem{comment} Strictly speaking the selection of nodes sending links to the
outside world is biased to those with higher values of $K_{out}$
and thus might be not representative of the whole community.
However, generally speaking this does not contradict our
mean-field approximation as (at least in random scale-free
networks) there is no correlation between the out-degree of a node
and its Google ranking.
%
%In the absence of other correlations that would mean that $G_c$ and $G_w$
%used in Eq. \ref{eq_1} should be calculated by weighting the appropriate set of
%nodes with values of their out-degrees:
%$G_{c}=\sum_{i \in C} G(i)  K_{out}(i)/\sum_{j \in C} K_{out}(j)$.
%
\bibitem{WWW_exponents}  A. Albert, H. Jeong, and A.-L. Barabási,
%Diameter of the World Wide Web,
Nature, {\bf 401}, 130,
%130-131
(1999).
%\bibitem{metropolis} N. Metropolis, et.al. J. Chem. Phys. {\bf 21},
%1087 (1953).
%
\bibitem{maslov_physicaA_2004} S. Maslov, K. Sneppen, A. Zaliznyak,
Physica A, {\bf 333}, 529 (2004).
%
\bibitem{newman_stub} M. E. J. Newman, S. H. Strogatz, and D. J.
Watts, Phys. Rev. E {\bf 64}, 161 (1995).
%
\bibitem{rewire} S. Maslov and K. Sneppen, Science, \textbf{296},
910 (2002).
\bibitem{newman_fermion}
J. Park and M. E. J. Newman,
%Origin of degree correlations in the Internet and other networks. (7 pages)
Phys. Rev. E {\bf 68}, 026112 (2003).
%
\end{thebibliography}
\end{document}